\documentclass[letterpaper,twocolumn,showpacs,prl,superscriptaddress,email]{revtex4-1} 

\usepackage[pdftex]{graphicx}  
\usepackage{amssymb}
\usepackage{amsmath,amsfonts,latexsym}
\usepackage{array,tabularx,color}
\usepackage{dcolumn}               
\usepackage[normalem]{ulem}
\usepackage{comment}
\usepackage{bm}
\usepackage[latin1]{inputenc}
\usepackage{color}

\newcommand{\vect}[1]{{\mathbf #1}}

\begin{document}

\title{Frictionless flow in a binary polariton superfluid}
\author{E.~Cancellieri}
\email[Corresponding author: ]{emiliano.cancellieri@uam.es}
\affiliation{F\'{\i}sica Te\'orica de la Materia Condensada,
Universidad Aut\'onoma de Madrid, Spain}
\author{F. M. Marchetti}
\affiliation{F\'{\i}sica Te\'orica de la Materia Condensada,
Universidad Aut\'onoma de Madrid, Spain}
\author{M. H. Szyma\'nska}
\affiliation{Department of Physics, University of Warwick,
Coventry, United Kingdom}
\author{D. Sanvitto}
\affiliation{NNL, Istituto Nanoscienze - CNR, Lecce, Italy}
\author{C. Tejedor}
\affiliation{F\'{\i}sica Te\'orica de la Materia Condensada,
Universidad Aut\'onoma de Madrid, Spain}

\date{\today}

\begin{abstract}
  We study the properties of a binary microcavity polariton superfluid
  coherently injected by two lasers at different momenta and
  energies. The crossover from the supersonic to the subsonic regime,
  where motion is frictionless, is described by evaluating the linear
  response of the system to a weak defect potential. We show that the
  coupling between the two components requires that either both
  components flow without friction or both scatter against the defect,
  though scattering can be small when the two fluids are weakly
  coupled. By analyzing the drag force exerted on a defect, we give a
  recipe to experimentally address the crossover from the supersonic
  to the subsonic regime.
\end{abstract}

\maketitle 

Coherent quantum fluids can undergo a transition to the superfluid
phase, where the fluid viscosity is zero. When the system excitations
are described in terms of quasi particles, the Landau
criterion~\cite{pitaevskiibook} establishes the value of the fluid
critical velocity below which no excitation can be created and the
fluid exhibits superfluidity. In particular, for weakly interacting
Bose-Einstein condensates (BECs), the critical velocity equals the
speed of sound. The description of the superfluid properties of
coupled multicomponent condensates, where each component can have a
different density, and so a different speed of sound, and a different
velocity, is far from trivial. Yet, exploring how the superfluid
properties of one fluid are modified by the presence of a second one
is of fundamental interest. Binary superfluids in cold atomic BECs
have recently attracted noticeable interest: here, the formation of
solitary waves (see, e.g., Ref.~\cite{berloff05}), the emission of
Cherenkov-like radiation from a dragged defect~\cite{susanto07}, and
the critical velocities~\cite{kravchenko08} have been studied.
Because of their versatility in control and detection, cavity
polaritons --- the strong coherent mixture of a quantum well exciton
with a cavity photon --- represent an ideal framework to address this
problem. In particular, the injection of polaritons by two external
laser fields allows us to independently tune the two fluid degrees of
freedom such as energies, momenta (and therefore flow velocities), and
particle densities, something not possible to implement in atomic
condensates. At the same time, their finite lifetime makes polaritons
prototypical systems for the study of condensation out of equilibrium.

Superfluidity in resonantly excited one component polariton fluids has
been tested both theoretically~\cite{carusotto2004,carusotto2005} and
experimentally~\cite{amo2009} through the observation of a dramatic
but not complete~\cite{cancellieri2010} reduction of the scattering
against a defect.
As far as multicomponent polariton fluids are concerned, superfluidity
has been demonstrated in the optical parametric oscillator (OPO)
regime through the frictionless propagation of wave
packets~\cite{amo2008} and the observation of quantized vortices and
persistent currents~\cite{sanvitto2010,marchetti10}. However, a
thorough analysis of the superfluid properties of multicurrent systems
is still missing.

In this Letter we consider a two-component polariton system resonantly
injected via two pumping lasers at different momenta and energies, and
analyze its superfluid properties. Following a Landau criterion
approach, we study the Bogoliubov excitation spectra in the linear
approximation, showing the conditions under which the system can
sustain frictionless flow, and analyzing how the superfluid properties
of one component depend on the density and velocity of the other
component.
We perform the linear response analysis for defects with size smaller
than the healing length. The case of bigger and stronger defects is
more complex since nonlinear waves can be emitted and a linear
analysis of the problem might not be sufficient~\cite{el2007}.
Remarkably, we find that, within the validity of the Landau criterion,
the possibility of the system to display frictionless flow in one
component and simultaneously a flow with friction in the other is
impeded by the coupling between the two components.
Naturally, when coupling a supersonic (SP) fluid with a subsonic (SB)
one, the amount of scattering induced by the SP component to the SB
one depends on the coupling strength between the two fluids and their
individual properties. Further, by making use of a full numerical
analysis of the system mean-field nonlinear dynamics, we study the
drag force exerted by both condensates on a defect, and give a recipe
to experimentally address the SB to supersonic SP crossover.

We describe the dynamics of resonantly-driven microcavity polaritons
via a Gross-Pitaevskii equation for coupled cavity ($\psi_C$) and
exciton ($\psi_X$) fields generalized to include decay and resonant
pumping ($\hbar=1$)~\cite{whittaker05}:
\begin{align}
  \nonumber i\partial_t \begin{pmatrix} \psi_X \\ \psi_C \end{pmatrix}
  &=
  \begin{pmatrix} 0 \\ F \end{pmatrix} + \left[\hat{H}_0
    + \begin{pmatrix} g_X|\psi_X|^2 & 0 \\ 0 &
      V_C \end{pmatrix}\right]
  \begin{pmatrix} \psi_X \\ \psi_C \end{pmatrix}\\
  \hat{H}_0 &= \begin{pmatrix} \omega_{X} - i \kappa_X & \Omega_R/2
    \\ \Omega_R/2 & \omega_{C}(-i\nabla) - i \kappa_C \end{pmatrix}\;
  .
\label{eq:model}
\end{align}
Here, two continuous wave pumping lasers, $F={\cal F}_1({\bf
  r})e^{i({\bf k}_1 \cdot{\bf r}-\omega_1t)}+{\cal F}_2({\bf
  r})e^{i({\bf k}_2\cdot{\bf r}- \omega_2t)}$ resonantly inject
polaritons at frequencies $\omega_{1,2}$ and momenta ${\bf k}_{1,2}$
-- both lasers pump along the $x$-axis, ${\bf k}_{1,2} =
(k_{1,2},0)$. We assume the exciton dispersion $\omega_X$ to be
constant and the cavity one quadratic, $\omega_C(-i\nabla)=\omega^0_C
-\frac{\nabla^2}{m_C}$, with $m_C=2\times10^{-5}m_0$ and $m_0$ being
the electron mass.  $\Omega_R$ is the Rabi frequency
($\Omega_R=4.4$~meV) and $\kappa_X$ and $\kappa_C$ are the excitonic
and photonic decay rates.  The exciton-exciton interaction strength
$g_X$ is set to one by rescaling both $\psi_{X,C}$ and ${\cal
  F}_{1,2}$. We set the energy zero to $\omega_X=\omega^0_C$ (zero
detuning). Finally, the potential $V_C(\vect{r})$ describes either a
defect naturally present in the cavity mirrors or generated by an
extra laser pump~\cite{amo2010}.

In the linear approximation regime, and for a homogeneous pump (${\cal
  F}_{1,2}({\bf r})=F_{1,2}$), we can limit our study to the following
approximated solution of the Gross-Pitaevskii equation
\begin{equation}
  \psi_{X,C}(\vect{r},t)=\sum_{j=1,2}e^{-i\omega_j t}
  \left[e^{i\vect{k}_j\cdot\vect{r}}
    \psi^{ss}_{j_{X,C}}+\theta_{j_{X,C}} (\vect{r,t})\right]\; ,
\label{eq:expa}
\end{equation}
where $\psi^{ss}_{j_{X,C}}$ are the mean-field steady state solutions,
and where $\theta_{i_{X,C}}(\vect{r},t)$ are small fluctuation fields
describing the linear response of the system to a weak defect
potential $V_C(\vect{r})$.
Similarly to Refs.~\cite{carusotto2004,carusotto2005}, by
substituting~\eqref{eq:expa} into~\eqref{eq:model}, at the zeroth
order ($\theta_{j_{X,C}} = 0 = V_C(\vect{r})$) the mean-field
solutions $\psi^{ss}_{j_{X,C}}$ solve a system of four coupled complex
equations, while the fluctuation fields as well as their
(Bogoliubov-like) spectra can be obtained by expanding linearly in
$\theta_{j_{X,C}}$ and $V_C(\vect{r})$. For additional details, see
the Supplemental Material and Ref.~\cite{cancellieri2011}.

\begin{figure}
  \centering
  \includegraphics[width=1.0\linewidth]{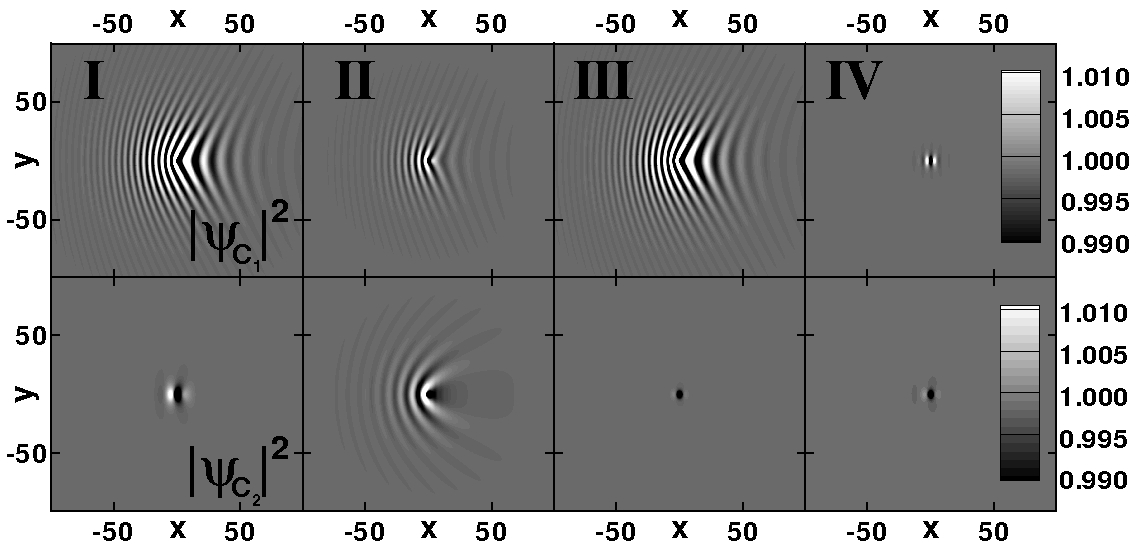}
  \includegraphics[width=1.0\linewidth]{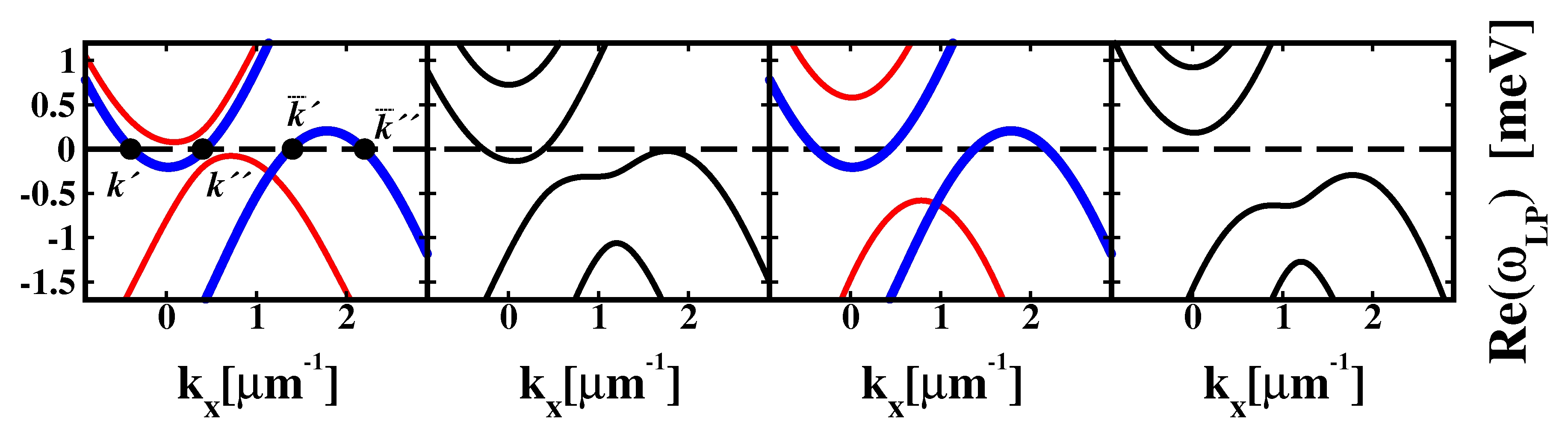}
  \includegraphics[width=1.0\linewidth]{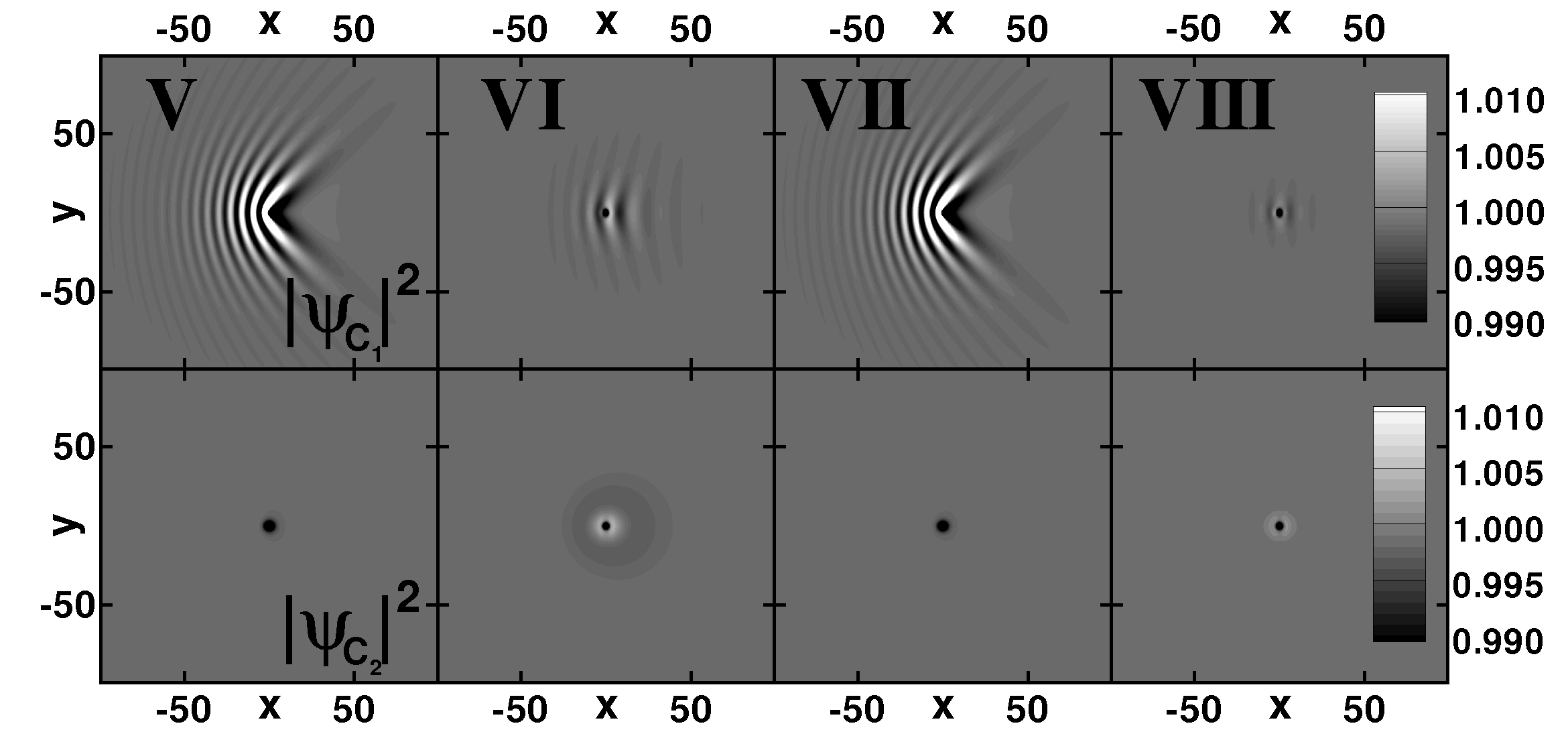}
  \includegraphics[width=1.0\linewidth]{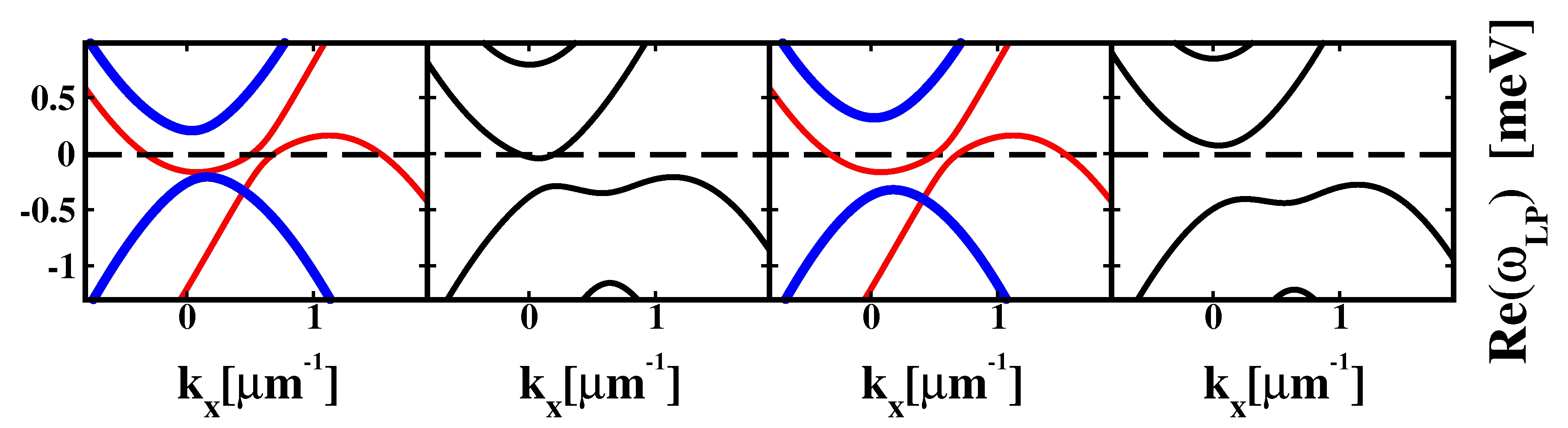}
  \caption{(Color online) 2D contour plots of the space profiles
    $|\psi_{{1,2}_C} (\vect{r})|^2$ [arbitrary units] (gray maps) and
    associated excitation spectra $\Re [ \omega_{LP} (\vect{k})]$
    [meV]: the two laser pumps are shined at momenta $k_{1}=0.9$ and
    $k_{2}=0.4~\mu$m$^{-1}$ (columns I-IV) and at momenta $k_{1}=0.6$
    and $k_{2}=0.1~\mu$m$^{-1}$ (columns V-VIII), while in all cases
    the laser energies are $0.5$~meV blue-detuned above the bare LP
    branch ($\kappa_C=\kappa_X=0.1$~meV). Columns I, III, V, and VII
    corresponds to the case where fluid 1 (red) is uncoupled from
    fluid 2 (blue), while columns II, IV, VI, and VIII describe the
    coupled cases. The densities of the two components have been fixed
    to $g_X |\psi_{1_X}|^2=1.5$~meV and $g_X|\psi_{2_X}|^2=1.2$~meV
    (columns I and II), to $g_X|\psi_{1_X}|^2=1.5$~meV and
    $g_X|\psi_{2_X}|^2=2.5$~meV (III and IV), to
    $g_X|\psi_{1_X}|^2=1.0$~meV and $g_X|\psi_{2_X}|^2=1.25$~meV (V
    and VI), and to $g_X|\psi_{1_X}|^2=1.0$~meV and
    $g_X|\psi_{2_X}|^2=1.5$~meV (VII and VIII). The momentum labels
    $k'=-0.4$, $k''=0.40$, $\bar{k}'=1.4$ and
    $\bar{k}''=2.2~\mu$m$^{-1}$ are explicitly indicated in the
    spectrum of column I.}
\label{fig:subsup}
\end{figure}
The SP vs. SB character of the excitations generated by the defect
potential can be studied by analyzing the real part of the Bogoliubov
spectra $\omega_{LP_j,UP_j}^{\pm}(\vect{k})$.
According to the Landau criterion for superfluidity, a fluid moving
against a defect is in a SB regime if it is unable to excite quasi
particle states (i.e., when elastic scattering is forbidden). This
happens when the system's excitation spectra, is either gapped, i.e.,
\begin{equation}
  \Re [ \omega_{LP_j}^{\pm}(\vect{k})] \ne 0 \quad \forall \vect{k}\,
  ,
\label{eq:crite}
\end{equation}
or it satisfies the condition $\Re[\omega_{LP_j}^{\pm}(\vect{k}_0)] =
0$ for one value of the momentum only, namely that of the condensate's
momentum $\vect{k}_0$ (linear spectrum). Conversely, when for at least
two values of $\vect{k}$, $\Re[\omega_{LP_j}^{\pm}(\vect{k})] = 0$,
the system is in the SP regime.
Note that, unlike for superfluid systems in thermal equilibrium, for
polaritons the above definition of the SB regime does not mean a
complete suppression of the energy dissipation into the creation of
quasi particles~\cite{wouters2010b,cancellieri2010}. In fact, because
of the polariton finite lifetime, the spectra are broadened and a
residual drag is always present.

In order to analyze the superfluid properties of the system, in
Fig.~\ref{fig:subsup} we compare the cases of coupled and uncoupled
fluids. This can be regarded, both from a theoretical and experimental
point of view, as the comparison between the case of two fluids pumped
in different regions of the cavity (uncoupled) with the case of two
fluids pumped in the same region (coupled). Clearly, the densities of
two coupled fluids depend on both pump intensities, and thus, in order
to correctly compare the coupled and uncoupled scenarios, such
intensities must be adjusted so that the polariton densities of each
fluid in the coupled case separately coincide with the ones of the
uncoupled fluids.
Typical behaviors of the system are illustrated in
Fig.~\ref{fig:subsup}, where both 2D contour plots of the space
profiles $|\psi_{{1,2}_C} (\vect{r})|^2$ and their associated
excitation spectra $\Re [ \omega_{LP} (\vect{k})]$ are plotted. Let us
consider first the case of the panels corresponding to columns I to
IV: for uncoupled components (columns I and III), the spectrum of
fluid $1$ (red) crosses the zero-energy line in four points at $k'$,
$k''$, $\bar{k}'$ and $\bar{k}''$, satisfying $k'+\bar{k}''=2k_{1}$
and $k''+\bar{k}'=2k_{1}$. Two quasi particles with momentum $k_{1}$
can be excited, and thus fluid $1$ is in the SP regime. Now
Cherenkov-like waves can be emitted from the $\delta$-like defect
positioned in $\vect{r}=0$ (see the $|\psi_{{1}_C} (\vect{r})|^2$ map
of column I).
In contrast, the spectrum of the fluid $2$ (blue) is gapped, no
Cherenkov waves can be emitted from the defect, and therefore fluid
$2$ is in the SB regime.
When, instead, we analyze the case where the same two fluids are
coupled (column II), we see that Cherenkov-like waves appear in the 2D
profiles of both $|\psi_{C_1}(\vect{r})|^2$ and
$|\psi_{C_2}(\vect{r})|^2$. This is because the interaction between
the two fluids produces an anticrossing, and thus a mixing, of the
corresponding Bogoliubov modes. As a consequence, the fluid injected
in the component $2$ can now scatter against the defect.
An opposite case is shown in columns III and IV. The polariton density
of fluid $2$ is now doubled with respect to the case of columns I and
II, keeping unchanged the fluid $1$ density. Now, the effect of the
coupling is to considerably decrease the scattering in component $1$
and the coupled excitation spectra satisfies Eq.~\eqref{eq:crite}: in
this case the effect of the coupling is that both components can flow
without friction.
From this analysis, we can conclude that a two-component polariton
fluid can be in SB regime only if both components are SB. This is
because, due to the coupling, the Bogoliubov spectra mix and only the
scattering properties of the system as a whole can be defined. Since
the combined state of the coupled system depends on the densities of
both fluids, the system as a whole is SP or SB depending on which
component dominates.
In addition, we find that when a fluid has either a too low density or
a too high velocity to exhibit frictionless flow on its own, the fluid
can instead flow without friction when coupled to another fluid with
the suitable properties.
In order to identify the role played by the coupling strength between
the two fluids in our predictions, we consider in columns V-VIII of
Fig.~\ref{fig:subsup} the case of two fluids with a higher photonic
component, and therefore more weakly coupled, with respect to the case
of columns I-IV. While the same qualitative conclusions hold, the
scattering induced by fluid $1$ over fluid $2$ is now substantially
smaller and comparable with the effect due to the polariton linewidth.
\begin{figure}
  \centering
  \includegraphics[width=1.0\linewidth]{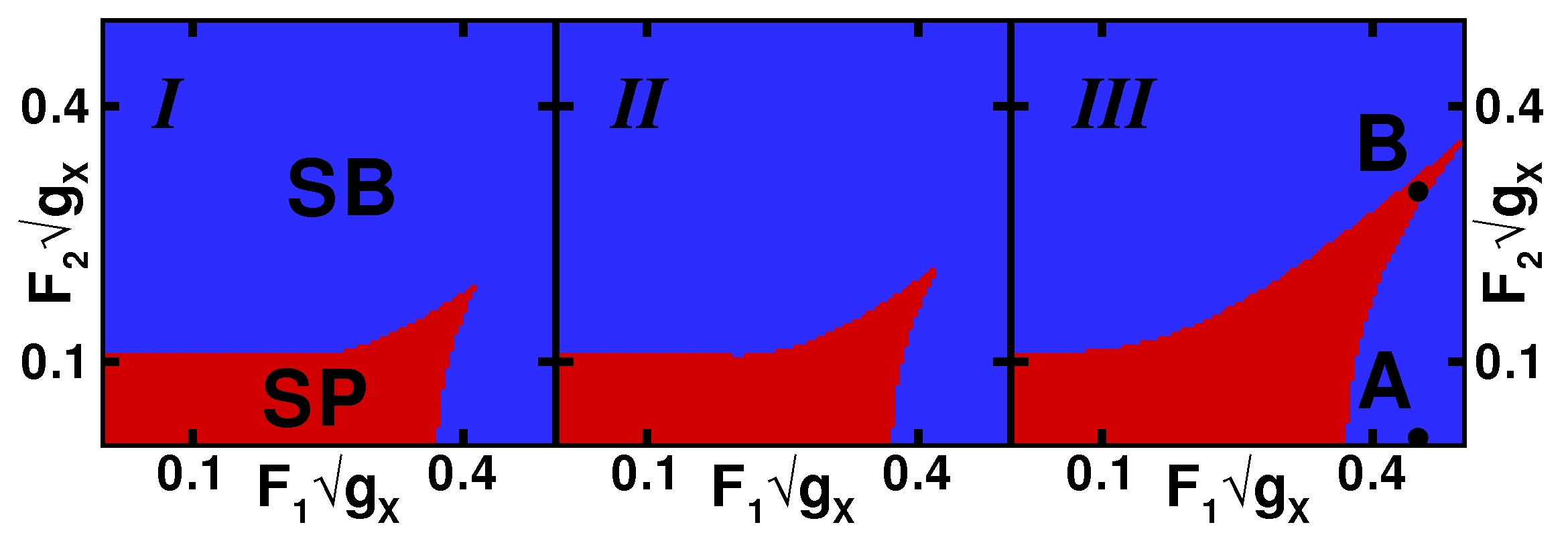}
  \caption{\label{fig:phasediagram}(Color online) Phase diagram, as a
    function of the rescaled pump intensities $\sqrt{g_X} F_{1,2}$
    [meV$^{3/2}$], showing the regions where the system is SP (red) or
    SB (blue). In this case, the two lasers pump at an energy
    $0.5$~meV blue-detuned from the bare LP branch and
    $\kappa_X=\kappa_C=0.1$~meV. In panel I, the lasers momenta are
    $k_{1}=0.6$, and $k_{2}=0.1~\mu$m$^{-1}$. In panels II and III the
    $x$-component of the momentum of laser 2 is increased by $0.2$ and
    $0.4~\mu$m$^{-1}$ respectively. Points $A$ and $B$ correspond to
    cases discussed in the text.}
\end{figure}
Applying Eq.~\eqref{eq:crite}, one can study the SP and SB character
of the binary fluid as a function of the two particle densities (see
Supplemental Material). It is however useful to perform this study
also as a function of the two pump intensities, being these the
experimentally accessible parameters. Panels I-III of
Fig.~\ref{fig:phasediagram} show the SB regions for three values of
the fluid $2$ velocity. Clearly, if any of the two pumps is switched
off ($F_j=0$), one reproduces the single-fluid case. As $k_{2}<k_{1}$,
the SB region for $F_1=0$ starts at lower pump intensities than the SB
region for $F_2=0$ (panel I): for higher fluid velocities, the system
requires higher polariton populations, and therefore higher pump
intensities, in order to be in the SB regime.
Even if the analytical dependence of the SB region on the two pump
intensities cannot be evaluated, one can qualitatively understand its
behavior: for fixed cavity and laser parameters, the SB regime depends
on the total particle density seen by the two components
$|\psi_X^{ss}|^2= |\psi_{1_X}^{ss}|^2+|\psi_{2_X}^{ss}|^2$. For
$F_2=0$ and $\sqrt{g_X} F_{1}=0.45$~meV$^{3/2}$ (point A of
Fig. \ref{fig:phasediagram}), the total particle density
$|\psi_X^{ss}|^2$, seen by the fluid is
$|\psi_X^{ss}|^2=|\psi_{1_X}|^2=1.37$~meV/$g_X$ and the system is
SB. If now the second pump is turned on and $\sqrt{g_X}F_2$ set to
$0.3$~meV$^{3/2}$ (point B), the total particle density decreases to
$|\psi_X^{ss}|^2=1.34$~meV/$g_X$ and the system is in the SP
regime. This is because when $F_2$ is turned on the particle density
increases by a factor $|\psi_{2_X}|^2$ but, at the same time, the
fluid $1$ particle density is decreased by a bigger factor. Since the
system starts in a SB regime, the dressed LP branch is blue-detuned
with respect to the pump frequency $\omega_1$ and, therefore, the
effect of $F_2\ne 0$ is to further bluedetune it, making it more
difficult for pump 1 to fill the cavity.

\begin{figure}
  \centering
  \includegraphics[width=1.0\linewidth]{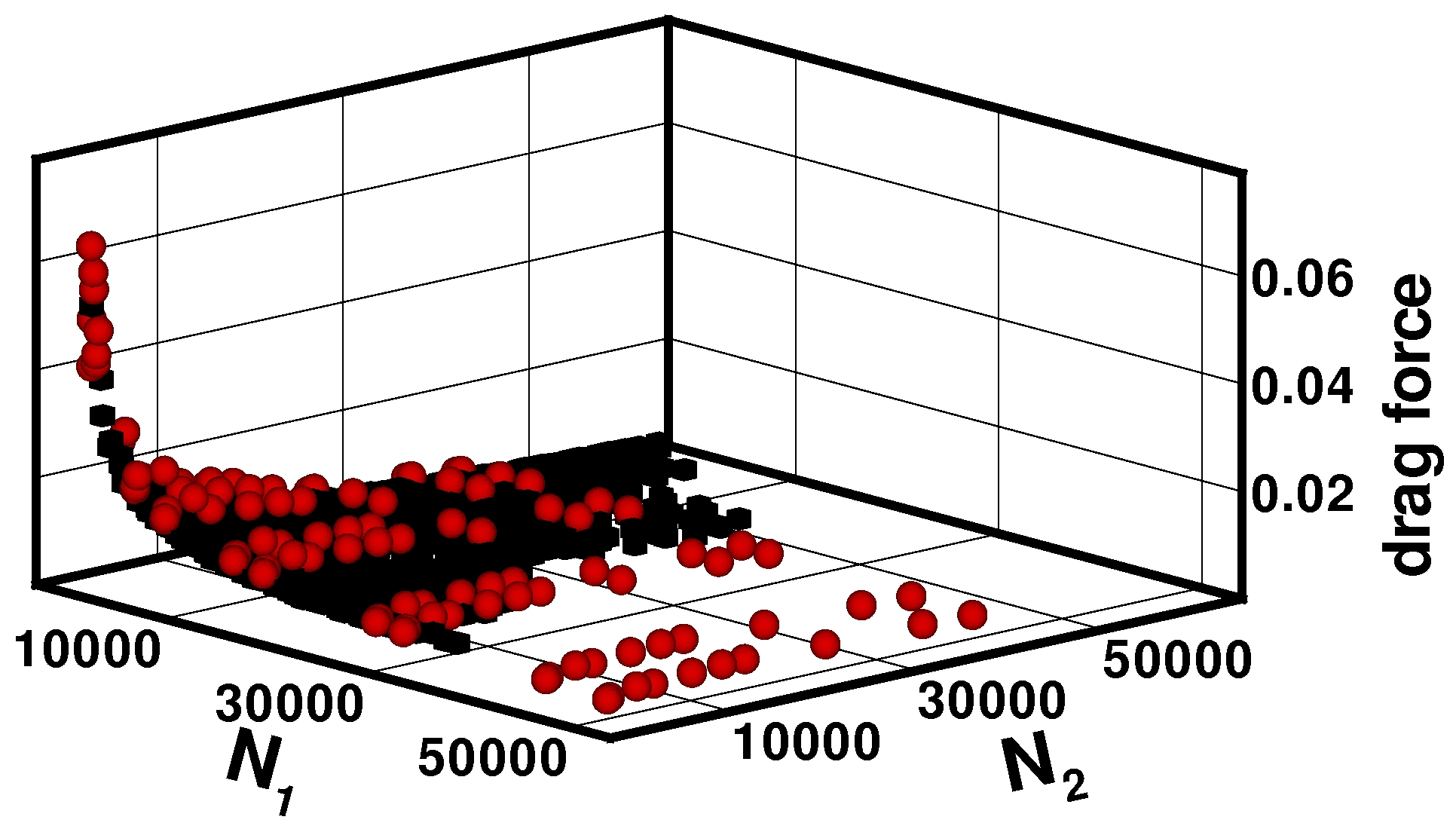}
  \caption{\label{fig:drag}(Color online) Time average of the drag
    force of the fluid as a function of the number of particles $N_1$
    and $N_2$. The two lasers are pumping at momenta $k_{1}=0.9$, and
    $k_{2}=1.0~\mu$m$^{-1}$ and energies $0.3$~meV blue-detuned from
    the bare LP branch, and $\kappa_X=0.22$~meV, $\kappa_C=0.22$~meV.
    Red dots correspond to the case of two uncoupled fluids, while
    black cubes correspond to the case of coupled fluids. The drag
    force for the uncoupled case is evaluated as: $\vect{F}_d =
    \frac{1}{N_1+N_2}(N_1\vect{F}_{d_1}+N_2\vect{F}_{d_2})$.}
\end{figure}

Evaluating the linear spectrum of excitations in experiments can be a
challenging task. In principle, the appearance and disappearance of
Cherenkov waves could be used to determine the SP to SB crossover,
similarly to Ref.~\cite{amo2009}. However, for a quantitative
description of the crossover we propose to determine the drag force
exerted by the binary fluid on the defect $V_C(\vect{r})$
\cite{wouters2010b,astrakharchik2004,cancellieri2010}:
\begin{equation}
  \vect{F}_d = \frac{1}{\int d\vect{r} |\psi_C(\vect{r})|^2}\int
  d\vect{r} |\psi_C(\vect{r})|^2 \nabla V_C (\vect{r})\; .
\label{eq:dragf}
\end{equation}
We evaluate the time average of the cavity field $\psi_C(\vect{r})$,
numerically solving the dynamics of Eq.~\eqref{eq:model} on a 2D grid
($256\times256$ points) of $150\times 150~\mu$m, by using a
fifth-order adaptive-step Runge-Kutta algorithm. The pumping lasers
have a smoothen top-hat spatial profile ${\cal F}_{1,2} ({\bf r})$
with a full width at half maximum of $\sigma=130~\mu$m; the weak
defect has a Gaussian shape.
In Fig.~\ref{fig:drag} we plot the drag force that the binary fluid
exerts on the defect as a function of the two fluid numbers of
particles, comparing the coupled and uncoupled cases. The limit when
one of the two pumps is turned off recovers the results for a single
fluid~\cite{cancellieri2010}: when the particle density increases, the
drag force decreases from high values to a residual finite value. For
the case with two currents, we find that the drag force exerted by two
coupled fluids on the defect is weaker than the drag force exerted by
the two uncoupled components. This is because, in the coupled case,
particles of each component move in an effectively denser medium than
in the uncoupled case (Eq. (3) of the Supplemental Material), thus the
drag force is smaller.
From the experimental point of view, in order to determine the drag
force, one could measure the near-field cavity emission in a region
around the defect as a function of position, and, if the shape of the
defect is known, one could evaluate the drag force making use of
Eq.~\eqref{eq:dragf}. Note, that the important quantity needed for
this measure is the shape of the potential, not its precise height.
Any uncertainty in the defect potential intensity will systematically
affect the drag force overall scale but not its global dependence on
the polariton densities. Finally, we would like to stress that higher
fluid velocities and shorter polariton lifetimes give rise to higher
values (therefore more easily measurable) of the drag force and of its
residual value at high polariton density.

To conclude, we would like to note that we can draw this Letter's
conclusions independently on the polariton lifetime, as they
exclusively depend on the real part of the Bogoliubov spectra and
therefore hold in equilibrium conditions, e.g., for the case of atomic
superfluids. However, even for extremely long polariton lifetimes,
binary polariton superfluids are more general than atomic ones. This
is because, while for the latter case the chemical potential fixes the
atom density, for the former, the laser frequency can be tuned
independently on its power which determines the polariton
density. Further, the polariton dispersion deviates from quadratic at
large momenta. This together with the finite polariton lifetime has
important consequences on the Bogoliubov spectrum, even in the case of
one fluid only: while for atoms, Bogoliubov spectra are all linear at
small wave vectors, for coherently pumped polaritons they can in
addition be gapped or diffusive~\cite{carusotto2005}.

This research has been supported by the Spanish MEC (MAT2008-01555,
QOIT-CSD2006-00019) and CAM (S-2009/ESP-1503). F.M.M. acknowledges
the financial support from the program Ram\'on y Cajal.

\newcommand\textdot{\.}

\newpage

\begin{widetext}

\begin{center}
\Large \textbf{Frictionless flow in a binary polariton superfluid:
  Supplementary Material}
\end{center}

\normalsize
\begin{center}
  This supplementary material contains, in Appendix A, the background
  information about the linear approximation solutions of the
  generalized Gross-Pitaevskii equation describing a binary
  microcavity polariton superfluid. We give the equations describing
  the mean-field steady state solutions of the system, and derive the
  linearized Bogoliubov-like theory, needed for the evaluation of the
  excitation spectra.  Moreover we give the equations for the
  evaluation of the response of the system to a weak external
  perturbation, and for the evaluation of the photonic and excitonic
  wave-functions in real space. In appendix B we introduce a phase
  diagram where the sub-sonic or super-sonic behavior of the
  suplefluid is studied as a function of the particle density.
\end{center}

\section{Appendix A}

\subsection{Stationary solutions in the homogeneous case}
We describe the system of resonantly-driven microcavity polaritons via
Gross-Pitaevskii equation for coupled cavity ($\psi_C$) and exciton
($\psi_X$) fields generalized to include the description of the finite
life-time of photons and excitons and the injection of polaritons in
the cavity through resonant pumping ($\hbar=1$)~\cite{whittaker05}:
\begin{equation}
  \nonumber i\partial_t \begin{pmatrix} \psi_X \\ \psi_C \end{pmatrix}
  =
  \begin{pmatrix} 0 \\ F \end{pmatrix} + \left[\hat{H}_0
    + \begin{pmatrix} g_X|\psi_X|^2 + V_X & 0 \\ 0 &
      V_C \end{pmatrix}\right]
  \begin{pmatrix} \psi_X \\ \psi_C \end{pmatrix}\\
\end{equation}
\begin{equation}
  \hat{H}_0 = \begin{pmatrix} \omega_{X} - i \kappa_X & \Omega_R/2
    \\ \Omega_R/2 & \omega_{C}(-i\nabla) - i \kappa_C \end{pmatrix}\;
  .
\label{eq:mode2}
\end{equation}
%
Polaritons are continuously injected into the cavity by two spatially
homogeneous continuous-wave laser fields:
\begin{equation*}
  F(\vect{r},t)=F_1 e^{i (\vect{k}_1 \cdot \vect{r} - \omega_1 t)}
  +F_2 e^{i (\vect{k}_2 \cdot \vect{r} - \omega_2 t)}\; ,
\end{equation*}
with intensities $F_{1,2}$, and with independently tunable frequencies
$\omega_{1,2}$ and momenta $\vect{k}_{1,2}$, which can be
experimentally changed by changing the laser angle of incidence with
respect to the growth direction. Under the continuous pump conditions
and in the homogeneous case (i.e., in absence of an external
potential, $V_{C,X}(\vect{r})=0$), the mean-field solutions of
Eq.~\eqref{eq:mode2} can be written as:
\begin{equation}
  \psi_{X,C}(\vect{r},t) =
  \psi^{ss}_{1_{X,C}}e^{i(\vect{k}_1\cdot\vect{r}-\omega_1 t)} +
  \psi^{ss}_{2_{X,C}}e^{i(\vect{k}_2\cdot\vect{r}-\omega_2 t)} \; .
\label{eq:meanf}
\end{equation}
Substituting the expression~\eqref{eq:meanf} into~\eqref{eq:mode2} we
obtain 4 contributions, two of which oscillate at the main frequencies
$\omega_1$ and $\omega_2$ and the additional two at the replica (or
satellite state) frequencies $\omega_1-\Delta\omega$ and
$\omega_2+\Delta\omega$, where $\Delta \omega=\omega_2-\omega_1$.
Similarly to what is done in the OPO
regime~\cite{whittaker05,wouters07} where replica states in addition
to the pump signal and idler states are neglected, here, we consider
only the terms oscillating at the main frequencies $\omega_1$ and
$\omega_2$. In this approximation the mean-field values for
$\psi^{ss}_{{1,2}_{X,C}}$, can be obtained solving the following
system of complex equations:
\begin{equation}
  \begin{cases}
  [\omega_X-\omega_1-i\kappa_X+G_{12}]\psi_{1_X}^{ss}+
  \frac{\Omega_R}{2}\psi_{1_C}^{ss}=0 \\
  [\omega_C({\bf k}_1)-\omega_1-i\kappa_C]\psi_{1_C}^{ss}+
  \frac{\Omega_R}{2}\psi_{1_X}^{ss}+F_1 =0 \\
  [\omega_X-\omega_2-i\kappa_X+G_{21}]\psi_{2_X}^{ss}+
  \frac{\Omega_R}{2}\psi_{2_C}^{ss}=0 \\
  [\omega_C({\bf k}_2)-\omega_2-i\kappa_C]\psi_{2_C}^{ss}+
  \frac{\Omega_R}{2}\psi_{2_X}^{ss} +F_2 =0 \; ,
\end{cases}
\label{eq:stead}
\end{equation}
where $G_{ij}=g_X(|\psi_{i_X}^{ss}|^2+2|\psi_{j_X}^{ss}|^2)$ with
$i\neq j=1, 2$. The mean-field system of equations~\eqref{eq:stead}
can have up to 9 solutions, i.e. 6 solutions more than in the case of
one pumping laser, but only a maximum of 3 solutions are stable. For
details see Ref.~\cite{cancellieri2011}.

\subsection{Linearized Bogoliubov-like theory}
The perturbative Bogoliubov-like analysis, first introduced for
resonantly pumped polaritons in
Refs.~\cite{carusotto2004,carusotto2005}, is here generalized to the
case of two pumping lasers. Adding small fluctuations to the
homogeneous solution, the Bogoliubov-like theory allows for the study
of the dynamical stability of the two-pump-frequency mean-field
solution, as well as for the study of the subsonic or supersonic
character of the fluid. Moreover, the Bogoliubov theory allows for the
evaluation of the real and momentum space representation of the
photonic and excitonic distributions in the presence of weak
perturbing potentials. We start our analysis by adding small
fluctuations ($\theta_{1,2_{X,C}}$) to the homogeneous solutions:
\begin{equation}
  \psi_{X,C}(\vect{r},t) = e^{-i\omega_1 t}
  \left[e^{i\vect{k}_1\cdot\vect{r}} \psi^{ss}_{1_{X,C}} +
    \theta_{1_{X,C}}(\vect{r,t})\right] + e^{-i\omega_2 t}
  \left[e^{i\vect{k}_2\cdot\vect{r}} \psi^{ss}_{2_{X,C}} +
    \theta_{2_{X,C}}(\vect{r,t})\right] \; ,
\label{eq:fluct}
\end{equation}
where the fluctuation fields can be divided into particle-like and
hole-like excitations:
\begin{equation*}
  \theta_{i_{X,C}}(\vect{r},t)=\sum_{\vect{k}} [e^{-i\omega
      t+i\vect{k}\cdot\vect{r}}u_{i_{X,C}\vect{k}}+e^{i\omega t+i(2
      \vect{k}_i-\vect{k})\cdot\vect{r}}v^*_{i_{X,C}\vect{k}}] \; .
\end{equation*}
Inserting Eq.~\eqref{eq:fluct} in Eq.~\eqref{eq:mode2} and expanding
up to linear terms in $\theta_{1,2_{X,C}}$, we obtain 4 terms
oscillating at frequencies $\omega_1-\Delta\omega\pm\omega$ and
$\omega_2+\Delta \omega\pm\omega$, which we neglect, and 4 terms
oscillating at $\omega_{1,2}\pm\omega$ which we consider. In other
words, we limit our study to the case where only the two states with
frequencies $\omega_{1,2}$ are occupied and analyze the excitation of
particles with frequencies $\omega_{1,2}\pm\omega$.

\subsubsection{Excitation Spectra}
Within this analysis, the stability of the system and the scattering
properties of the collective excitations are evaluated through the
spectra of the particle-like $u_{i_{X,C}\vect{k}}$ and of the
hole-like $v^*_{i_{X,C}\vect{k}}$ excitations which are given by the
solutions of the eigenvalue equation:
\begin{equation}
  \left[\omega \mathbb{I} -
    \mathbb{L}_{\vect{k}}\right]\mathbb{U}_{\vect{k}} = \left[\omega
    \mathbb{I} - \begin{pmatrix} \mathbb{L}_{11 \vect{k}}&
      \mathbb{L}_{12 \vect{k}} \\ \mathbb{L}_{21 \vect{k}} &
      \mathbb{L}_{22 \vect{k}} \end{pmatrix}\right]
  \mathbb{U}_{\vect{k}} = 0\; ,
\label{eq:eigen}
\end{equation}
where the excitation fields have been arranged in the 8-component
vector
$\mathbb{U}^{\text{T}}=(u_{1_X},u_{1_C},v_{1_X},v_{1_C},u_{2_X},u_{2_C},v_{2_X},v_{2_C})$. In
the above equation the matrices $\mathbb{L}_{ij \vect{k}}$ with $i\ne
j$ are given by
\begin{equation*}
  2g_X e^{i(\vect{k}_i - \vect{k}_j)\cdot \vect{r}} \begin{pmatrix}
    \psi_{i_X}^{ss}\psi_{j_X}^{ss\star} & 0 & \psi_{i_X}^{ss}
    \psi_{j_X}^{ss} & 0 \\ 0 & 0 & 0 & 0 \\ -
    \psi_{i_X}^{ss\star}\psi_{j_X}^{ss\star} & 0 & -
    \psi_{i_X}^{ss\star} \psi_{j_X}^{ss} & 0 \\ 0 & 0 & 0 & 0
  \end{pmatrix}
\end{equation*}
and $\mathbb{L}_{jj \vect{k}}$ are given by
\begin{equation*}
  \begin{pmatrix}
  \omega_X - \omega_j - i\kappa_X + 2g_X|\psi_X^{ss}|^2 &
  \frac{\Omega_R}{2} & g_X\psi_{j_X}^{ss}\psi_{j_X}^{ss} & 0
  \\ \frac{\Omega_R}{2} & \omega_C(\vect{k}) - \omega_j - i\kappa_C &
  0 & 0 \\ -g_X \psi_{j_X}^{ss\star} \psi_{j_X}^{ss\star} & 0 &
  -\omega_X (2 \vect{k}_j- \vect{k}) + \omega_j - i\kappa_X -
  2g_X|\psi_X^{ss}|^2 & -\frac{\Omega_R}{2} \\ 0 & 0 &
  -\frac{\Omega_R}{2} & - \omega_C (2 \vect{k}_j- \vect{k}) + \omega_j
  - i\kappa_C \end{pmatrix}\; ,
\end{equation*}
with $|\psi_X^{ss}|^2=|\psi_{1_X}^{ss}|^2+|\psi_{2_X}^{ss}|^2$ being
the total excitonic density. At given values of the pumping strength
$F_1$ and $F_2$, the solutions of the mean-field
equations~\eqref{eq:stead} are stable if all the eight eigenvalues
$\omega_{LP_j,UP_j}^{\pm}(\vect{k})$ of Eq.~\eqref{eq:eigen} have
negative imaginary part for every value of the momentum $\vect{k}$.
When the stability of the solution for given values of the pump
intensities has been checked, the information about the scattering
properties of the fluctuations can be extracted by the analysis of the
real part of the eight eigenvalues.

\subsubsection{Response to a weak potential}
We evaluate now the response of the excitonic and photonic density
profiles to a weak static potential $V_{C}(\vect{r})$. The starting
point is the equation of motion for the fluctuation fields in the
presence of a perturbation~\cite{carusotto2005}:
$i\partial_t\mathbb{U}_{\vect{k}}=\mathbb{L}_{\vect{k}}\mathbb{U}_{\vect{k}}
+\mathbb{P}_{\vect{k}}$, where
\begin{equation}
  \mathbb{P}_{\vect{k}}=
  \begin{pmatrix}

\tilde{V}_X(\vect{k})\psi_{1_X}^{ss}\\ \tilde{V}_C(\vect{k})\psi_{1_C}^{ss}\\ -\tilde{V}_X
(\vect{k}-2\vect{k}_1)\psi_{1_X}^{*ss}\\ -\tilde{V}_C(\vect{k}-2\vect{k}_1)\psi_{1_C}^{*ss}\\ \tilde{V}_X(\vect{k})\psi_{2_X}^{ss}\\ \tilde{V}_C(\vect{k})\psi_{2_C}^{ss}\\ -\tilde{V}_X(\vect{k}-2\vect{k}_2)\psi_{2_X}^{*ss}\\ -\tilde{V}_C(\vect{k}-2\vect{k}_2)\psi_{2_C}^{*ss}\\
\end{pmatrix}\; ,
\end{equation}
where $\tilde{V}_{C,X}(\vect{k})$ is the Fourier transform into
momentum space of $V_{C,X}(\vect{r})$. Since we are interested in the
steady state of the system we can extract the perturbed photonic and
excitonic fields in momentum space as:
\begin{equation}
  \mathbb{U}_{\vect{k}}=-\mathbb{L}^{-1}_{\vect{k}}\mathbb{P}_{\vect{k}}\;
  ,
\end{equation}
and back-Fourier transform them in order to obtain the perturbation fields
in real space. At this point, the total photon/exciton field intensity
for each component (homogeneous solution + potential induced perturbation),
normalized to the intensity of the homogeneous solution without the potential
is obtained as:
\begin{equation}
|\psi_{i_{C,X}}|^2=\frac{|\psi^{ss}_{i_{C,X}}+\theta_{i_{X,C}}(\vect{r},t)|^2}
{|\psi^{ss}_{i_{C,X}}|^2}\; .
\end{equation}
\begin{figure}
  \centering
  \includegraphics[width=1.0\linewidth]{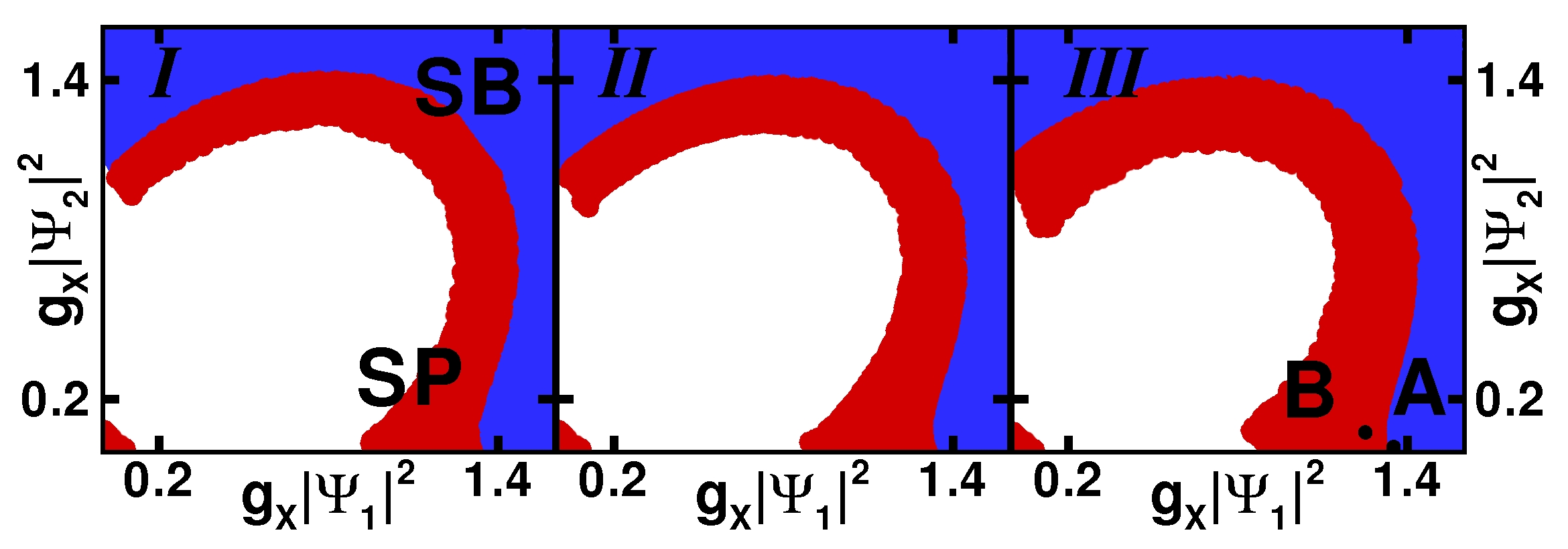}
  \caption{(Color online) Phase diagram, as a
    function of $g_X|\psi_{i_X}|^2$ [meV], showing the regions where
    the system is SP (red) or SB (blue).  The white regions correspond
    do particle densities for which the system is not stable. In this
    case, the two lasers pump at an energy $0.5$~meV blue detuned from
    the bare LP branch and $\kappa_X=\kappa_C=0.1$~meV.  In Panel $I$,
    the lasers momenta are $k_{1}=0.6$, and $k_{2}=0.1~\mu$m$^{-1}$.
    In Panels $II$ and $III$ the $x$-component of the momentum of
    laser 2 is increased by $0.2$ and $0.4~\mu$m$^{-1}$
    respectively. The points $A$ and $B$ correspond to cases discussed
    in the text.}
\label{fig:phasediagra2}
\end{figure}

\section{Appendix B}

\subsection{Phase diagram for the sub (super) sonic behavior of a binary fluid}
Looking at the spectra of small excitations over the stationary state,
it is possible to determine the sub-sonic (SB) or super-sonic (SP)
behavior of the coupled binary fluid. For given values of the lasers
and cavity parameters, it is possible to evaluate the stability of the
system and, for the stable conditions, the existence of a SB
solution. As discussed in the article, if the real part of the
excitation spectra satisfies the condition:
\begin{equation} 
  \Re [ \omega_{LP_j}^{\pm}(\vect{k})] \ne 0 \quad \forall \vect{k}\,
  ,
\end{equation}
the fluid is considered to be SB. In Fig.~\ref{fig:phasediagra2} we
plot the stable solutions that are SB (blue) or SP (red) as a function
of the particle densities of the two modes $|\psi_{i_X}|^2$. The three
panels reproduce the same laser and cavity conditions as panels
$I-III$ of Fig. 2 of the article. In the case of one pump turned off
(for example $F_2$, i.e. $|\psi_{2_X}|^2=0$) we have a SP region at
low densities, a wide region for which the system is not stable (in
white), a second SP region and, finally, a SB one that starts at
$g_X|\psi_{1_X}|^2=1.36$.  In Panel $III$ we show the two points $A$
and $B$ as in the article. Point $A$ is in the SB region while point
$B$ is in the SP one. Note, that the boundary between the SB and the
SP region is not perfectly circular meaning that the SB behavior
depends on the two densities but in an unbalanced way.  This can be
understood recalling that polaritons at different energies and momenta
have different masses and couplings, and therefore their weight on the
SB behavior is different.

\end{widetext}

\end{document}